\documentstyle[aps,preprint,prd]{revtex}
\date{\today}
\begin{document}
\flushbottom

\widetext
\draft
\title{Coulomb potential from a particle in uniform ultrarelativistic motion}
\author{A. J. Baltz}
\address{
Physics Department,
Brookhaven National Laboratory,
Upton, New York 11973}
\date{\today}
\maketitle

\def\thepage{\arabic{page}}
\makeatletter
\global\@specialpagefalse
\ifnum\c@page=1
\def\@oddhead{Draft\hfill To be submitted to Phys. Rev. C}
\else
\def\@oddhead{\hfill}
\fi
\let\@evenhead\@oddhead
\def\@oddfoot{\reset@font\rm\hfill \thepage \hfill}
\let\@evenfoot\@oddfoot
\makeatother

\begin{abstract}
The Coulomb potential produced by an ultrarelativistic particle (such as a
heavy ion) in uniform motion is shown in the appropriate gauge to factorize
into a longitudinal $\mbox{\boldmath $\delta$}(z - t)$
dependence times the simple two dimensional potential
solution in the transverse direction.  This form makes manifest the source of
the energy independence of the interaction.
\end{abstract}

\makeatletter
\global\@specialpagefalse
\def\@oddhead{\hfill}
\let\@evenhead\@oddhead
\makeatother
\nopagebreak
\narrowtext
In the consideration of the electromagnetic interaction of ultrarelativistic
heavy ions, to produce for example electron-positron pairs, one may assume
uniform straight line trajectories for the moving ions and then express
the time-dependent classical effect as the Li\'enard-Wiechert potential
\begin{equation}
V(\mbox{\boldmath $ \rho$},z,t)={\alpha Z(1-v\alpha_z)\over
\sqrt{ [({\bf b}-\mbox{\boldmath $ \rho$})/\gamma]^2+(z-v t)^2}}.
\end{equation}
${\bf b}$ is the impact parameter, perpendicular to the $z$--axis along which
the ion travels, $\mbox{\boldmath $\rho$}$, $z$, and $t$ are the coordinates of
the potential relative to a fixed target (or ion)
$\alpha_z$ is the Dirac matrix,
and $Z, v$ and $\gamma$ are the charge, velocity and $\gamma$ factor the
moving ion $(\gamma=1/\sqrt{1-v^2})$.  It has previously been shown that
in the large $\gamma$ limit
a gauge transformation on this potential can remove both the large
positive and negative time contributions of the potential as well as the
$\gamma$ dependence \cite{brw}.  If one makes the gauge
transformation on the wave function
\begin{equation}
\psi=e^{-i\chi({\bf r},t)} \psi'
\end{equation}
where
\begin{equation}
\chi({\bf r},t)={\alpha Z \over v} \ln [\gamma(z-v t)+\sqrt{b^2+\gamma^2(z-v
t)^2}]
\end{equation}
the interaction $V({\bf\rho},z,t)$ is gauge transformed to
\begin{equation}
V(\mbox{\boldmath $ \rho$},z,t)={\alpha Z(1-v\alpha_z)\over
\sqrt{ [({\bf b}-\mbox{\boldmath $ \rho$})/\gamma]^2+(z-v t)^2}}-{\alpha
Z(1-(1/v)\alpha_z)\over\sqrt{b^2/\gamma^2+(z-v t)^2}}.
\end{equation}
The work of Toshima and Eichler\cite{te} previouly utilized a similar transform
to solve the difficulty in calculations with interactions that drop off
slowly as $1 / t$.

In the limit of large $\gamma$, impact parameter not too large, $v=1$,
and Eq.(4) can be expressed in a simple multipole
decomposition\cite{brw},
\begin{eqnarray}
V(\mbox{\boldmath $ \rho$},z,t)&=&\alpha Z(1-\alpha_z) { 2 \pi \over r}
\nonumber \\
&& \times \biggl\{-\ln {r^2 +t^2 \over b^2} \sum_l  Y_l^0(\theta_t,0)
Y_l^0(\theta,0)\ \ \ \ r>\sqrt{b^2+t^2} \nonumber \\
&&+\sum_{m>0}{2 \cos m \phi \over m}
\sum_l Y_l^m(\theta_t,0) Y_l^m(\theta,0)\nonumber \\
&&\times\biggl[\biggl({r^2-t^2 \over b^2}\biggr)^{m/2}
\ \ \ \ \ \ \ \ \ \ \ \ \  |t|<r<\sqrt{b^2+t^2}\nonumber\\
&&+\biggl({b^2 \over r^2 - t^2}\biggr)^{m/2}\biggr]\biggr\},
\ \ \ \ \ \ \ \ \ \ \ \ \   r>\sqrt{b^2+t^2}
\end{eqnarray}
where $\cos \theta_t = t/r$, and $V(\mbox{\boldmath $ \rho$},z,t) = 0$ in the
region $r < |t|$.
If one notices that for any $m$
\begin{equation}
\sum_l Y_l^m(\theta_t,0) Y_l^m(\theta,0) = {1 \over 2 \pi}
\mbox{\boldmath $\delta$}(\cos\theta - \cos\theta_t)
\end{equation}
then the expression becomes
\begin{eqnarray}
V(\mbox{\boldmath $ \rho$},z,t)&=&\alpha Z(1-\alpha_z)
\mbox{\boldmath $\delta$}(z - t) \nonumber \\
&& \biggl\{ -\ln {\rho^2 \over b^2}\ \ \ \ \ \ \ \ \ \ \ \ \  \rho>b\nonumber\\
&&+\sum_{m>0}{2 \cos m \phi \over m}\nonumber\\
&&\times\biggl[\biggl({\rho \over b}\biggr)^m
 \ \ \ \ \ \rho <b \nonumber\\
&&+\biggl({b \over \rho}\biggr)^m\biggr]\biggr\}.
 \ \ \ \ \ \rho >b
\end{eqnarray}
The expression now has cylindrical symmetry,
with a delta function in $(z-t)$ and a Fourier series in $\phi$ which involves
only $\rho$ and $b$ in the coefficients.

The expression of Eq.(5) was obtained by ignoring terms of order
$\ln \gamma / \gamma^2 $\cite{brw}.  This means that to the same accuracy in
$ 1 / \gamma$ the $(z-t)$ dependence of the potential can be adequately
represented by $\mbox{\boldmath $\delta$}(z - t)$ in
Eq.(7).  We would like to find the function of the transverse variable
$V(\mbox{\boldmath $ \rho$})$ whose Fourier series is expressed in Eq.(7)
\begin{equation}
V(\mbox{\boldmath $ \rho$},z,t)
=V(\mbox{\boldmath $ \rho$})
\mbox{\boldmath $\delta$}(z - t).
\end{equation}
$V(\mbox{\boldmath $ \rho$})$ may be evaluated from the form of Eq. (4):
\begin{eqnarray}
V(\mbox{\boldmath $ \rho$})
&=& \int_{-\infty}^{\infty}
V(\mbox{\boldmath $ \rho$},z,t) d z\nonumber \\
&=& \alpha Z(1-\alpha_z) \int_{-\infty}^{\infty}\biggl[{1 \over
\sqrt{ [({\bf b}-\mbox{\boldmath $ \rho$})/\gamma]^2+(z- t)^2}}-{ 1
\over\sqrt{b^2/\gamma^2+(z- t)^2}}\biggr] d z \nonumber \\
&=& \alpha Z(1-\alpha_z) \ln { z' + \sqrt{z'^2 + [({\bf
b}-\mbox{\boldmath$\rho$})/\gamma]^2} \over z' + \sqrt{z'^2 + b^2 /
\gamma^2}}\Bigg|_{-\infty}^{\infty}.
\end{eqnarray}
The gauge transformed form allows the removal of the infinities and factors of
$\gamma$ in the evaluation of the integral.  The result is
\begin{equation}
V(\mbox{\boldmath $ \rho$},z,t)
=-\mbox{\boldmath $\delta$}(z - t) \alpha Z(1-\alpha_z)
\ln{({\bf b}-\mbox{\boldmath $ \rho$})^2 \over b^2}.
\end{equation}
One can construct the Fourier series for Eq.(10) and verify that the result is
identical to Eq.(7).

The $b^2$ in the denominator of the logarithm is removable by a gauge
transformation.  If one wished to have a potential with the same gauge
for all impact parameters one would remove it.  The $b^2$ has no effect
on the $m > 0$ terms in Eq.(7) but just adds the constant seen in the $m = 0$
term.

Of course, once the existence of the effective two dimensional potential
is established, one really doesn't have to carry out the integral of Eq.(9);
the solution
\begin{equation}
V(\mbox{\boldmath $ \rho$})
=-\alpha Z(1-\alpha_z)
\ln\ ({\bf b}-\mbox{\boldmath $ \rho$})^2.
\end{equation}
immediately follows from Maxwell's equations and symmetry.

This note has pointed out that implicit in the previously presented multipole
decomposition formula, Eq.(5), is the extremely simple cylindrically
symmetrical expression, Eq.(10).  These expressions are valid from the
smallest relevant impact parameters just outside nuclear touching, and
throughout the region where $b$ is not too large and the interaction is strong
enough to be non-perturbative.  At large enough $b / \gamma$, the
$\mbox{\boldmath $\delta$}(z - t)$ form becomes a poor approximation, but then
alternative perturbation theory expressions can be utilized for calculating
processes like electron-positron pair production\cite{brwp}

Unfortunately, the result of Eq.(10) has no obvious application to
coupled channels calculations utilizing the conventional angular momentum
algebra\cite{rum} \cite{brwb} \cite{brwf}, since radial multipole forms such
as those of Eq.(5) are necessary
when angular integration is done analytically.  However, the reduction of
the interaction from three dimensions to the two of Eq.(10) might make direct
solution of the time-dependent Dirac equation, without using coupled channels,
a viable alternative for the calculation of pair production induced by
ultrarelativistic heavy ions.
\vskip .5cm
This manuscript has been authored under Contract No. DE-AC02-76-CH00016 with
the U. S. Department of Energy

\end{document}